\documentclass[epj]{svjour}
\usepackage{graphics}
\usepackage{epsfig}
\usepackage[numbers,sort&compress]{natbib}
\usepackage[utf8]{inputenc}
\usepackage{amsmath}
\usepackage{amssymb}
\usepackage[colorlinks=true
,urlcolor=blue
,anchorcolor=blue
,citecolor=blue
,filecolor=blue
,linkcolor=blue
,menucolor=blue
,pagecolor=blue
,linktocpage=true
,pdfproducer=medialab
]{hyperref}
\begin{document}
\title{Interpretations of galactic center gamma-ray excess confronting the PandaX-II constraints on dark matter-neutron spin-dependent scatterings in the NMSSM}

\author{Liangliang Shang\inst{1} \and Yangle He\inst{1} \and Jingwei Lian\inst{1} \and Yusi Pan\inst{1}
}
\institute{$^1$Department of Physics,
                Henan Normal University, Xinxiang 453007, China}
\date{Received: date / Revised version: date}
%
\abstract{
The Weakly Interacting Massive Particle (WIMP) has been one of the most attractive candidates for Dark Matter (DM), and the lightest neutralino ($\widetilde{\chi}^0_1$) in the Next-to-Minimal Supersymmetric Standard Model (NMSSM) is an interesting realization of the WIMP framework. The Galactic Center Excess (GCE) indicated from the analysis of the photon data of the \textit{Fermi} Large Area Telescope (Fermi-LAT) in the gamma-ray wavelength $\lesssim 1 \,{\rm fm}$, can be explained by WIMP DM annihilations in the sky, as shown in many existing works.
In this work we consider an interesting scenario
in the $Z_3$-NMSSM where the singlet $S$ and Singlino $\widetilde{S}^0$ components play important roles in the Higgs and DM sector.
Guided by our analytical arguments, we perform a sophisticated scan over the NMSSM parameter space
by considering various observables such as the Standard Model (SM) Higgs data measured by the ATLAS and CMS experiments at the Large Hadron Collider (LHC), and the $B$-physics observables $BR(B_s\rightarrow X_s\gamma)$ and $BR(B_s\rightarrow \mu^+\mu^-)$.
We first collect samples which can explain the GCE well while passing all constraints we consider except for the DM direct detection (DD) bounds from XENON1T and PandaX-II experiments. We analyze the features of these samples suitable for the GCE interpretation and find that $\widetilde{\chi}^0_1$ DM are mostly Singlino-like and annihilation products are mostly the bottom quark pairs $\bar{b}b$ through a light singlet-like CP-odd Higgs $A_1$. Moreover, a good fit to the GCE spectrum generically requires sizable DM annihilation rates $\langle \sigma_{b\bar{b}} v \rangle _{0}$ in today's Universe. However, the correlation between the coupling $C_{A_1 b\bar{b}}$ in $\langle \sigma_{b\bar{b}} v \rangle _{0}$ and the coupling $C_{Z \widetilde{\chi}^0_1 \widetilde{\chi}^0_1}$ in DM-neutron Spin Dependent (SD) scattering rate $\sigma^{SD}_{\widetilde{\chi}^0_1-N}$ makes all samples we obtain for GCE explanation get excluded by the PandaX-II results. Although the DM resonant annihilation scenarios may be beyond the reach of our analytical approximations and scan strategy, the aforementioned correlation can be a reasonable motivation for future experiments such as PandaX-nT to further test the NMSSM interpretation of GCE.
} 
\maketitle
\section{Introduction}
\label{intro}
An excess of gamma-rays in the direction of the Galactic Center (GCE) has been reported by several groups analyzing the data from the Large Area Telescope on board the \textit{Fermi} Gamma-ray Space Telescope (Fermi-LAT) \cite{TheFermi-LAT:2015kwa,Daylan:2014rsa,Calore:2014xka,Hooper:2013rwa,Gordon:2013vta,Abazajian:2012pn,Hooper:2011ti,Hooper:2010mq,Goodenough:2009gk,Zhou:2014lva,Huang:2015rlu,Fermi-LAT:2017vmf}.
Various interpretations have been proposed to provide additional gamma-ray sources which can be roughly classified into two categories. One is of astrophysical origin such as a large population of unresolved millisecond pulsars (MSPs) in the Galactic bulge \cite{Fermi-LAT:2017yoi,Cholis:2014lta,Lee:2015fea,Bartels:2015aea,Petrovic:2014xra,Hooper:2013nhl,Hooper:2015jlu,Hooper:2016rap,Brandt:2015ula}
or a series of recent leptonic cosmic-ray outbursts \cite{Cholis:2015dea,Petrovic:2014uda,Carlson:2014cwa}. 
Another category is of particle Dark Matter (DM) origin which annihilate into the Standard Model (SM) particles subsequently producing gamma-rays in the excess energy range (see e.g.~\cite{Ipek:2014gua,Boehm:2014hva, Hooper:2014fda,Berlin:2014pya,Agrawal:2014una,Berlin:2014tja,Izaguirre:2014vva,Cheung:2014lqa,Cerdeno:2014cda,Alves:2014yha,Hooper:2012cw,Ko:2014gha,Boehm:2014bia,Abdullah:2014lla,Martin:2014sxa,Cline:2014dwa,Kim:2016csm,Karwin:2016tsw,Ghorbani:2014qpa,Abazajian:2014fta,Cao:2014efa,Cao:2015loa,Gherghetta:2015ysa} and references therein).
 For the comparison between the two categories of GCE interpretation, it has been argued that \cite{Haggard:2017lyq} a sufficiently large population of MSPs would  imply a large population of observable low-mass X-ray binaries, limiting the contribution of MSPs to the GCE to $\sim 4\%-23\%$ and thus leaving annihilating DM as a contender. Another argument comes from \cite{Hooper:2013nhl} by comparing the GCE spectrum to that measured from 37 MSPs by \textit{Fermi} experiment, which indicates that a population of unresolved MSPs exhibit a spectral shape that is too soft at sub-GeV energies to accommodate the GCE. Furthermore, \cite{Hooper:2016rap} argued that after pulsars are expelled from a globular cluster, they continue to lose rotational kinetic energy and become less luminous. This makes the luminosity function of those MSPs depart from the steady-state distribution and thus MSPs born in globular clusters can account for only a few percent or less of the GCE.
Consequently, pulsars located in the Galactic bulge or a Galactic halo of DM are both viable interpretations for the GCE.

The existence of DM has been verified by various cosmological and astrophysical observations. However, no direct evidence of DM particle interactions has been measured by any experiment. DM particle search is a promising strategy to find new particle physics principles Beyond the Standard Model (BSM). Weakly Interacting Massive Particle (WIMP) has been a popular DM candidate given that its scatterings off the SM nuclei mediated by the weak interaction may be detected \cite{Goodman:1984dc,Feng:2010gw}.
One of the most attractive WIMPs is the Lightest Supersymmetric Particle (LSP) in supersymmetric models (SUSY) such as the lightest neutralino ($\widetilde{\chi}^0_1$).

Among the various realizations of the SUSY framework, the Minimal Supersymmetric Standard Model (MSSM) is the simplest one and provides a promising WIMP DM. However, after a series of experiments upgraded their sensitivities and set stronger bounds, MSSM is no longer a favorable framework for WIMP. Firstly, the SM-like Higgs mass measured to be around 125 GeV by the ATLAS and CMS experiment at the Large Hadron Collider (LHC) \cite{ATLAS:2012tfa,CMS:2012xdj} requires a sizable loop radiative corrections through heavy top squark ($\gtrsim1$ TeV) which leads to a large fine tuning.
Secondly, to generate the observed DM relic density while surviving the DM direct detection limits simultaneously, the Bino ($\widetilde{B}^0$) must be the dominant ingredient in the lightest neutralino. Consequently, the $\mu$ parameter in MSSM have to be relatively large which further aggravates the fine tuning problem.

As a minimal extension of the MSSM, the Next-to-MSSM (NMSSM) contains an additional gauge singlet superfield $\hat{S}$ which can dynamically generate the $\mu$ parameter in MSSM with a small value, in which case the fine tuning problem can be alleviated significantly \cite{Cao:2012fz}.
Recently, several experiments including LUX \cite{LUX:2016vxi}, PandaX \cite{PandaX:2016ega} and XENON \cite{Aprile:2017iyp} updated their results of DM direct detections (DD).
At this moment, the strongest constraints on Spin-Independent (SI) and Spin-Dependent (SD) WIMP-nucleon scattering cross sections come from XENON1T ($\sigma^{SI}_{\widetilde{\chi}^0_1\text{-}P}$) and PandaX-II ($\sigma^{SD}_{\widetilde{\chi}^0_1\text{-}N}$), respectively.
These stringent limits have already pushed the neutralino WIMP DM scenario in NMSSM into a challenging situation: one either imposes large cancelations in the DM-nucleon interactions among different contributions, the so-called blind spot scenario \cite{Cheung:2012qy,Huang:2014xua,Badziak:2015exr,Badziak:2016qwg,Cao:2016cnv,Baum:2017enm}, or chooses a large value of $\mu$ parameter as did in the MSSM \cite{Cao:2016cnv}. Either choice makes the NMSSM also suffer from the difficulty of fine tuning.

In \cite{Cao:2015loa} we observed that a light neutralino $\widetilde\chi_1^0$ in the $Z_3$-NMSSM was able to explain the GCE meanwhile satisfying various other constraints including the observed DM relic density and the Higgs data measured by the ATLAS and CMS groups at the LHC, by adjusting the $\widetilde\chi_1^0$ pair annihilation branch ratios into $b\bar{b}$, $W^+W^-$ and $A_1H_i$ ($i=1,2$).
These DM annihilations can generate additional gamma-ray sources strong enough to interpret the observed GCE and present a close connection between the astrophysical gamma-ray observations and DM phenomenology. More interestingly, the LSP $\widetilde\chi_1^0$ usually manifests itself as large missing transverse momentum at high energy colliders such as the LHC, which is one of the most important signals for searching SUSY.

In this work we present our dedicated analysis on a correlation pattern between the DM annihilation cross sections to explain the GCE and DM-nucleon spin-dependent scattering rates in the direct detection experiments,
in the $Z_3$-NMSSM where the singlet $S$ and Singlino $\widetilde{S}^0$ components play important roles in the Higgs and DM sector.
Guided by the analytical arguments, we perform a sophisticated scan by employing the Markov Chain Monte Carlo (MCMC) strategy over the NMSSM parameter space, to verify our expectations and explore whether numerical calculations can reveal NMSSM parameter space that can still explain the GCE while passing the limits from XENON1T and PandaX-II.

We organize the paper as follows. In Section \ref{sec.connection} we recapitulate the main features of neutralino DM in the NMSSM and the DM annihilation mechanisms to explain the GCE. Using the analytical approximations of spin-dependent DM-nucleon scattering rate, we identify the correlation between the GCE interpretation and the scattering strength. In Section \ref{sec.strategies} we explain our scan strategies in detail. We present and analyze the results of our scan in Section \ref{sec.result} and summarize in Section \ref{sec.summary}.

\section{Neutralino DM in the NMSSM }
\label{sec.connection}

In this section, we briefly recapitulate the features of neutralino DM in NMSSM and the mechanisms to explain the GCE in terms of DM annihilations. Then we discuss the analytical approximations of neutralino DM-nucleon spin-dependent scattering rate, based on which we identify the correlation between the GCE interpretation and the scattering strength in direct detection.

\subsection{Lightest neutralino in NMSSM as WIMP DM}
The NMSSM introduces a gauge singlet Higgs superfield $\hat{S}$ in additional to the two doublet Higgs superfield $\hat{H_u},\hat{H_d}$ in MSSM. A $Z_3$ symmetry is adopted to the superpotential of NMSSM to forbid the appearance of parameters with mass dimension. In this case, the superpotential of NMSSM is given by \cite{Ellwanger:2009dp}:
\begin{eqnarray}
  W_{\rm NMSSM} &=& W_F + \lambda\hat{H_u} \cdot \hat{H_d} \hat{S}
  +\frac{1}{3}\kappa \hat{S^3},
\end{eqnarray}
where $W_F$ is the superpotential of MSSM without the $\mu$-term and $\lambda$, $\kappa$ are dimensionless parameters that describe the interactions among the NMSSM Higgs superfields.
The soft SUSY-breaking Higgs potential is:
\begin{eqnarray}
V_{\rm NMSSM}^{\rm soft} &=& m_{H_u}^2 |H_u|^2 + m_{H_d}^2|H_d|^2
  + m_{H_S}^2|S|^2 \cr
  & &+( \lambda A_{\lambda} SH_u\cdot H_d
  +\frac{1}{3}\kappa A_{\kappa} S^3 + h.c.), \label{input-parameter1}
\end{eqnarray}
where $H_u$, $H_d$ and $S$ are the scalar component fields of $\hat{H}_u$,  $\hat{H}_d$ and $\hat{S}$, respectively, while $m_{H_u}$, $m_{H_d}$, $m_{H_S}$, $A_\lambda$ and $A_\kappa$ are soft SUSY-breaking parameters.
After the electroweak symmetry breaking, $m_Z$, $\tan \beta=v_u/v_d$ and $\mu_{eff}=\lambda v_s $ can be chosen as input parameters instead of $m_{H_u}^2$, $m_{H_d}^2$ and $m_{H_s}^2$, where
$v_u$, $v_d$, $v_s$ are vacuum expectation values (vev) of $H_u$, $H_d$ and $S$.
Moreover, we replace $A_\lambda$ by $M_A$ which is the mass of the would-be CP-odd Higgs in MSSM, using the following relation:
\begin{equation}
\label{eq:mA-doublet}
M_A^2=\frac{2\mu}{\sin{2\beta}}(A_\lambda+\kappa v_s).
\end{equation}
As a result, the Higgs sector of NMSSM has six free input parameters:
\begin{equation}
\lambda,\ \kappa,\ \tan{\beta},\ \mu,\ M_A,\ A_\kappa.
\end{equation}

NMSSM neutralino sector is obtained from the mixings among five gauginos: Bino $\widetilde{B}^0$, Wino $\widetilde{W}^0$, Higgsinos $\widetilde{H}_{d,u}^0$ and Singlino $\widetilde{S}^0$. The symmetric neutralino mass matrix in the basis of $(\psi^0)^T=(-i\widetilde{B}^0,-i\widetilde{W}^0,\widetilde{H}^0_d,\widetilde{H}^0_u,\widetilde{S}^0)$ can be written as
\begin{equation}
{\cal M} = \left(
\begin{array}{ccccc}
M_1 & 0 & -\frac{g_1 v_d}{\sqrt{2}} & \frac{g_1 v_u}{\sqrt{2}} & 0 \\
  & M_2 & \frac{g_2 v_d}{\sqrt{2}} & - \frac{g_2 v_u}{\sqrt{2}} &0 \\
& & 0 & -\mu & -\lambda v_u \\
& & & 0 & -\lambda v_d\\
& & & & \frac{2 \kappa}{\lambda} \mu
\end{array}
\right), \label{eq:MN}
\end{equation}
where $M_1$, $M_2$ are soft SUSY-breaking masses of Bino and Wino, respectively. After diagonalization one obtains five mass eigenstates $\widetilde{\chi}^0_i=N_{ij}\psi^0_j$ with ascending order of mass values, where $N_{ij}$ is the rotation matrix. The lightest neutralino $\widetilde{\chi}^0_1$ is identified as the WIMP DM in our discussion, which can be written as
\begin{eqnarray}
\label{eq:neutralino-decomposition}
\widetilde{\chi}^0_1 &=& N_{11} (-i \widetilde{B}^0)+N_{12} (-i \widetilde{W}^0) \nonumber \\
&+& N_{13} \widetilde{H}^0_d + N_{14} \widetilde{H}^0_u + N_{15} \widetilde{S}^0.
\end{eqnarray}

\subsection{Explaining GCE with NMSSM neutralino DM}
The main goal of identifying the lightest neutralino in NMSSM as WIMP DM is to use their annihilations in today's Universe to provide additional gamma-ray source in the Galactic Center (GC) region to explain the GCE. Therefore an adequate cross section of present neutralino annihilations is necessary. One can express the relevant annihilation processes as
\begin{equation}
\label{eq:channel-general}
\widetilde{\chi}^0_1 \widetilde{\chi}^0_1 \rightarrow XY.
\end{equation}
In \cite{Cao:2015loa} we observed that the relevant final states $XY$ can be bottom quark pair $b\bar{b}$, $W$ boson pair $W^+W^-$, and two scalars $A_1H_{1,2}$ where $H_{1,2}$ and $A_1$ are CP-even and CP-odd Higgs bosons, respectively. These SM particles as DM annihilation products will further generate photon spectrum including the gamma-ray energy range through decay, parton showering and hadronization effects \cite{Cirelli:2010xx}. Taking the $XY=b\bar{b}$ in \cite{Cao:2015loa} as an example, the DM annihilation cross section $\langle\sigma_{b\bar{b}} v\rangle_0$ in today's Universe with $v\to 0$ can be approximated as \cite{PhysRevD.41.3565}
\begin{equation}
  \langle \sigma_{b\bar{b}} v \rangle _{0}  \thickapprox
  \frac{3 \pi}{2} \sum_{i=1}^2 \frac{C_{A_i \widetilde{\chi}_1^0 \widetilde{\chi}_1^0}^2 C_{A_i b\bar{b}}^2 m_{\widetilde{\chi}_1^0}^2}{(4m_{\widetilde{\chi}_1^0}^2-m_{A_i}^2)^2+m_{A_i}^2\Gamma_{A_i}^2},
  \label{eq.bbar}
\end{equation}
where $C_{A_i \widetilde{\chi}_1^0 \widetilde{\chi}_1^0}$ and $C_{A_i b\bar{b}}$ are couplings. $A_i$ with $i=1,2$ are the CP-odd Higgs and $\Gamma_{A_i}$ are the widths of $A_i$.
We learned from \cite{Cao:2015loa} that the $s$-channel $A_i$-mediations dominate the $\widetilde{\chi}^0_1$ pair annihilation into $b\bar{b}$.

The predicted gamma-ray spectrum from DM annihilations in our Galaxy that are observed near Earth can be expressed as the following differential flux for a given angular direction:
\begin{eqnarray}
    \frac{ \mathrm{d}\Phi_{\gamma}}{ \mathrm{d}E_{\gamma} \mathrm{d}\Omega}
    &=&
    \frac{r_\odot \rho_\odot^2}{8\pi m_{\widetilde{\chi}_1^0}^2} J
    \sum\limits_{XY} \langle \sigma_{XY} v \rangle_0 \frac{ \mathrm{d}N^{XY}_{\gamma}}{\mathrm{d}E_\gamma},
    \\
    J &=& \int\limits_{l.o.s.}\! \frac{\mathrm{d}s}{r_\odot}\left(\frac{\rho(r(s,\theta))}{\rho_\odot}\right)^2,
    \label{Jfactor}
    \\
    \rho(r)&=&\rho_\odot \left(\frac{r}{r_\odot}\right)^{-\gamma}\left(\frac{1+r_\odot/R_s}{1+r/R_s}\right)^{3-\gamma},
    \label{eq.NFW}
\end{eqnarray}
where $dN^{XY}_\gamma /dE_\gamma$ is the prompt gamma-ray spectrum from a single DM annihilation event into $XY$.
$J$ is an astrophysical factor reflecting the Galactic distribution of DM obtained by integrating over $\rho(r)^2$ along the line of sight (l.o.s.) with $\rho(r)$ being the Navarro-Frenk-White (NFW) DM profile \cite{Navarro:1995iw}. We choose the following parameters in $\rho(r)$ as the same ones chosen in the Fermi-LAT analyses \cite{Fermi-LAT:2017vmf,Ackermann:2015lka-1506.00013}, which are consistent with various astrophysical observations (
e.g. the measured gravitational potential of the inner Galaxy \cite{Goodenough:2009gk}) while generally providing a good fit to the observed GCE \cite{Calore:2014xka,Abazajian:2014fta}.
\begin{equation}
\begin{array}{ccll}
r_\odot &=&8.5\ (\mathrm{kpc})& \mathrm{distance\ from\ sun\ to\ GC}\\
\rho_\odot &=&0.4 \ (\mathrm{GeV/cm^3})& \mathrm{DM\ density\ at\ location\ of\ Sun}\\
R_s &=&20\ (\mathrm{kpc})& \mathrm{typical\ scale\ radius} \\
\gamma&=&1.26 & r(s,\theta) = (r_\odot^2+s^2-2r_\odot s \cos{\theta})^{\frac{1}{2}}
\end{array}
\end{equation}
where $s$ is the l.o.s distance and $\theta$ is the angle between the l.o.s and the axis connecting Sun and GC.

\subsection{Correlation of GCE explanation to neutralino DM-neutron SD scattering in NMSSM}
\label{section-GCE-DD-analytical}

$XY=b\bar{b}$ in Eq.(\ref{eq:channel-general}) has been indicated in many existing works as a characteristic channel of DM annihilation to interpret GCE.
Eq.(\ref{eq.bbar}) shows that the corresponding DM annihilation rate is approximately proportional to the coupling $C_{A_1b\bar{b}}$ in NMSSM with a generally heavy $A_2$ \cite{Cao:2015loa,Ellwanger:2009dp},
\begin{eqnarray}
\label{eq:CA1bb}
C_{A_1b\bar{b}} = y_b \left( P_{1,A}\tan\beta \right) \equiv y_b R_{C_{A_1 b\bar{b}}},
\end{eqnarray}
where $y_b$ is the SM bottom quark Yukawa coupling and $|P_{1,A}|^2$ is the ingredient in the NMSSM CP-odd Higgs $A_1$ from the would-be CP-odd Higgs $A$ in MSSM with mass given in Eq.(\ref{eq:mA-doublet}). For a light singlet-like $A_1$, one usually has $|P_{1,A}|^2 \lesssim \mathcal{O}(0.5)$. In the case of relatively large $\tan\beta \gtrsim \mathcal{O}(5)$ which is common in the NMSSM \cite{Ellwanger:2009dp} and suitable for GCE explanation \cite{Cao:2015loa}, we can have the following approximation
\begin{eqnarray}
\langle \sigma_{b\bar{b}} v \rangle _{0} \propto \tan^2\beta,
\end{eqnarray}
as long as $P_{1,A}$ is small but stable for a light singlet-like $A_1$, which is the case we consider.

In \cite{Cao:2015loa} we observed the potential of future DM direct detection experiments to test the GCE explanation scenario in NMSSM, especially in terms of the DM-neutron spin dependent scattering rate $\sigma^{SD}_{\widetilde{\chi}^0_1-N}$ (see Figure 7 therein).
The dominant contribution comes from the $t$-channel mediation by $Z$ boson
\begin{eqnarray}
\mathcal{L}^{SD} = \sum\limits_{q=u,d} C_{\widetilde{\chi}^0_1 q} (\overline{\widetilde{\chi}^0_1} \gamma^\mu \gamma^5 \widetilde{\chi}^0_1) ( \bar{q} \gamma_\mu \gamma_5 q),
\end{eqnarray}
where the coupling $C_{\widetilde{\chi}^0_1 q}$ can be approximated in the limit of non-relativistic scattering as follows
\begin{eqnarray}
C_{\widetilde{\chi}^0_1 q} = \frac{C_{Z \widetilde{\chi}^0_1 \widetilde{\chi}^0_1} \, C_{Z\bar{q}q}}{m_Z^2}, \quad C_{Z \widetilde{\chi}^0_1 \widetilde{\chi}^0_1} = \frac{m_Z}{\sqrt{2}v}  \left(|N_{13}|^2 - |N_{14}|^2 \right),\nonumber
\end{eqnarray}
where $v\simeq 174$ GeV is the SM Higgs vev and $N_{ij}$ is rotation matrix for neutralino mass diagnolization in Eq.(\ref{eq:neutralino-decomposition}). $N_{ij}$ can have simple approximated expressions when DM $\widetilde{\chi}^0_1$ is dominated by one or two components (see Appendix of \cite{Cao:2015loa}). For example, when Bino $\widetilde{B}^0$ and Wino $\widetilde{W}^0$ are heavily decoupled while only Higgsinos $\widetilde{H}^0_d, \widetilde{H}^0_u$ and Singlino $\widetilde{S}^0$ are light, one has
\begin{eqnarray}
\frac{1}{|N_{15}|^2} \Big( |N_{13}|^2 - |N_{14}|^2 \Big) &\approx& \frac{\lambda^2 v^2}{m_{\widetilde{\chi}^0_1}^2-\mu^2} (1-\frac{1}{\tan^2\beta}),\\
N_{15} &=& (1+\frac{N_{13}^2}{N_{15}^2} +\frac{N_{14}^2}{N_{15}^2})^{-\frac{1}{2}},
\end{eqnarray}
where $|N_{15}|^2 \gtrsim \mathcal{O}(0.5)$ applies to the Singlino-like $\widetilde{\chi}^0_1$, as the case in most of our samples.
One can see that $\tan\beta\neq 1$ breaks the symmetric structure with respect to the Higgsinos $\widetilde{H}^0_{d},\widetilde{H}^0_{u}$ in the neutralino mass matrix Eq.(\ref{eq:MN}) and generates non-zero coupling between $Z$ boson and $\widetilde{\chi}^0_1$. More importantly, $\tan\beta>1$ in NMSSM can increase $|C_{Z \widetilde{\chi}^0_1 \widetilde{\chi}^0_1}|$ and the resulting DM-neutron SD scattering rate \cite{Belanger:2008sj}
\begin{eqnarray}
\sigma^{SD}_{\widetilde{\chi}^0_1-N} = \frac{16 \mu_r^2}{\pi} \frac{J_N+1}{J_N} \sum\limits_{q=u,d} |C_{\widetilde{\chi}^0_1 q}|^2 |S_N^q|^2,
\end{eqnarray}
where $q=u,d$ are the valence quarks inside neutron $N$ and $J_N=1/2$ is the neutron spin. $\mu_r=m_{\widetilde{\chi}^0_1}m_N/(m_{\widetilde{\chi}^0_1}+m_N)$ is the reduced mass between DM $\widetilde{\chi}^0_1$ and neutron $N$, which can be approximated to be $\mu_r \approx m_N \approx 1$ GeV for $m_{\widetilde{\chi}^0_1}\gtrsim \mathcal{O}(10)$ GeV as indicated in most existing GCE studies (also in this work). $S^q_N$ is the spin contribution of parton $q$ in neutron $N$.
Therefore in the simplified case of Higgsino-Singlino mixing, we have
\begin{eqnarray}
\sigma^{SD}_{\widetilde{\chi}^0_1-N} \propto |C_{Z \widetilde{\chi}^0_1 \widetilde{\chi}^0_1}|^2 \propto (1-\frac{1}{\tan^2\beta})^2.
\end{eqnarray}
In most samples we study in this work, the Singlino component dominates in $\widetilde{\chi}^0_1$.
However, we note that if additional components such as Bino $\widetilde{B}^0$ and Wino $\widetilde{W}^0$ also contribute sizably to $\widetilde{\chi}^0_1$, the above approximation may not apply very well.

To summarize this subsection, we obtained an approximately quantitative connections in terms of sizable $\tan\beta \gtrsim \mathcal{O}(5)$ in NMSSM with a singlet-like $A_1$ and a Singlino-like $\widetilde{\chi}^0_1$, between the GCE explanation and DM-neutron spin-dependent scattering rate as follows
\begin{eqnarray}
\label{eq:correlation}
\left\{
  \begin{aligned}
  &\langle \sigma_{b\bar{b}} v \rangle _{0} \propto R^2_{C_{A_1 b\bar{b}}}  \propto   \tan^2\beta. \\
  &\sigma^{SD}_{\widetilde{\chi}^0_1-N} \propto \Big( |N_{13}|^2 - |N_{14}|^2 \Big)^2 \propto  (1-\frac{1}{\tan^2\beta})^2, \\
  \end{aligned}
\right.
\end{eqnarray}

\section{Scan strategies}
\label{sec.strategies}
The scan over the NMSSM parameter space was performed by employing the MCMC method. We utilize NMSSMTools-5.0 \cite{Ellwanger:2004xm,Ellwanger:2005dv} to calculate the NMSSM mass spectrums, particle decaying ratios and $B$-physics observables, which are further interfaced to MicrOMEGAs \cite{Belanger:2008sj,Barducci:2016pcb} to calculate the DM annihilation cross section, relic density, DM-nucleon scattering rates, and the gamma-ray spectrum to compared with GCE. We build an overall $\chi^2$ function with which smaller values reflect better compatibility of a model sample in the NMSSM parameter space with the observables contained in $\chi^2$. The MCMC scan strategy is powerful in finding the NMSSM parameter regions meeting our physical goals.

\subsection{NMSSM parameter simplifications and ranges}
In our scan, NMSSM parameters not closely relevant to the GCE discussion are chosen with simplifications.
The first and second generation squarks are required to decouple with masses fixed at $2$~TeV.
For up- and down-type squarks of the third generation, we simplify the soft-SUSY breaking parameters to be $A_t=A_b$ and $M_{U_3}=M_{D_3}$.
All parameters in the slepton sector are chosen to be equal, including their masses and soft-SUSY breaking couplings. The gluino mass parameter $M_3$ is also fixed at 2~TeV. To summarize, the ranges of free NMSSM parameters in the scan are listed below:
\begin{equation}
\begin{array}{l}
2<\tan{\beta}<60,\quad 0.001<\lambda<1,\quad 0.001<\kappa<1,\\
100\,\mathrm{GeV}<\mu<1.5\,\mathrm{TeV},\quad 20\,\mathrm{GeV}<M_1<1\,\mathrm{TeV},\\
100\,\mathrm{GeV}<M_2<2\,\mathrm{TeV},\quad |A_{top}|<6\,\mathrm{TeV}, \\
100\,\mathrm{GeV}<M_{Q_3}<2\,\mathrm{TeV},\quad 100\,\mathrm{GeV}<M_{U_3}<2\,\mathrm{TeV},\\
100\,\mathrm{GeV}<M_{L}=M_{e}=A_e<2\,\mathrm{TeV}.
\end{array}
\end{equation}

\subsection{$\chi^2$ method in MCMC strategy}
Now we briefly recapitulate the $\chi^2$ method based on which we embed our codes of MCMC chain into NMSSMTools-5.0. Smaller $\chi^2$ of a NMSSM sample can be regarded as its better compatibility confronting the experimental observations built into the $\chi^2$. Smaller $\chi^2$ also result in higher acceptance probability of this sample which will act as the new starting point to search nearby parameter space better fitting the experiments. Note that a model point may still be accepted as a node in the scanning chain even if it predicts one or several observables outside the experimental bounds at some confidence level, as long as its $\chi^2$ is not very large. This strategy can help the codes avoid to get stuck in some corners of the total parameter space. When $\chi^2$ reaches a stable minimum after a long scan chain, we will apply a set of selection rules with respect to the relevant observables and name the kept points as \textit{GCE-samples}.

In this work we construct an overall $\chi^2_{total}$ in Eq.(\ref{eq.X2_total}) which consists of the SM-like Higgs mass, $B$-physics observables, DM relic abundance, DM-nucleon scattering rates, and the shape of GCE spectrum.
\begin{eqnarray}
\begin{array}{ccl}
\chi^2_{total} & = &\sum\limits_{i} \mathcal{F}_i \chi^2_i , \\
\chi^2_i & = & \left\{
  \begin{aligned}
  &\chi^2_{m_{h_{SM}}}, \chi^2_{BR(B_s \rightarrow X_s\gamma)}, \chi^2_{BR(B_s\rightarrow \mu^+\mu^-)}, \chi^2_{\Omega_{\widetilde{\chi}^0_1} h^2}  \\
   &\chi^2_{\sigma^{SI}_{\widetilde{\chi}^0_1-P}}, \chi^2_{\sigma^{SD}_{\widetilde{\chi}^0_1-N}} \\
   &\chi^2_{GCE}.\\
  \end{aligned}
\right.
\end{array}
\label{eq.X2_total}
\end{eqnarray}
In the above, $\mathcal{F}_i,m_{h_{SM}},BR,\Omega_{\widetilde{\chi}^0_1} h^2$ are weight factors, the SM-like Higgs mass, branching ratio and DM relic abundance, respectively. During the scan we dynamically adjust $\mathcal{F}_i$ to explore as wide regions as possible of the NMSSM parameter space, otherwise $\chi^2$ of one observable may overwhelm others' effects. $\sigma^{SI}_{\widetilde{\chi}^0_1-P}$ and $\sigma^{SD}_{\widetilde{\chi}^0_1-N}$ stand for the Spin-Independent (SI) DM-proton and Spin-Dependent DM-neutron scattering cross sections, respectively.

Some details about Eq.(\ref{eq.X2_total}) are as follows:

\begin{itemize}
\item The first type of $\chi^2$ in Eq.(\ref{eq.X2_total}), i.e. $\chi^2_{m_{h_{SM}}}$, $\chi^2_{BR(B_s \rightarrow X_s\gamma)}$, $\chi^2_{BR(B_s \rightarrow \mu^+\mu^-)}$ and $\chi^2_{\Omega_{\widetilde{\chi}^0_1} h^2}$, correspond to the case in which an observable has a theoretical prediction $\mu_i$, an observed center value $\mu_0$ and uncertainty $\sigma_i$ (including both experimental and theoretical parts). In this case the $\chi^2_i$ is defined as
\begin{equation}
\chi^2_i=\left(\frac{\mu_i-\mu_0}{\sigma_i}\right)^2,
\label{chi2-1D}
\end{equation}
in which the values of $\mu_i,\mu_0,\sigma_i$ will be provided when we apply the final-step selection cuts.

\item The second type of $\chi^2$ in Eq.(\ref{eq.X2_total}) is built based on the upper bounds of DM-nucleon scattering cross sections, due to the absence of confirmed signals in DM direct detections. We utilize the forms of $\chi^2_{\sigma^{SI}_{\widetilde{\chi}^0_1-P}}$ and $\chi^2_{\sigma^{SD}_{\widetilde{\chi}^0_1-N}}$ according to \cite{Matsumoto:2016hbs}:
\begin{eqnarray}
    \chi^2_{\sigma^{SI}_{\widetilde{\chi}^0_1-P}}&=&\frac{1}{2}\left(\frac{\sigma^{SI}_{\widetilde{\chi}^0_1-P}}{\delta\sigma^{SI}_{\widetilde{\chi}^0_1-P}}\right)^2,\\
    \chi^2_{\sigma^{SD}_{\widetilde{\chi}^0_1-N}}&=&\frac{1}{2}\left(\frac{\sigma^{SD}_{\widetilde{\chi}^0_1-N}}{\delta\sigma^{SD}_{\widetilde{\chi}^0_1-N}}\right)^2,
\end{eqnarray}
with
\begin{eqnarray}
    \delta\sigma^{SI}_{\widetilde{\chi}^0_1-P}&=&\left(\sigma^{SI,0}_{\widetilde{\chi}^0_1-P}/1.64\right)^2+(0.2\sigma^{SI}_{\widetilde{\chi}^0_1-P})^2,\\
    \delta\sigma^{SD}_{\widetilde{\chi}^0_1-N}&=&\left(\sigma^{SD,0}_{\widetilde{\chi}^0_1-N}/1.64\right)^2+(0.2\sigma^{SD}_{\widetilde{\chi}^0_1-N})^2,
\end{eqnarray}
where $\sigma^{SI,0}_{\widetilde{\chi}^0_1-P}$ and $\sigma^{SD,0}_{\widetilde{\chi}^0_1-N}$ are upper limits of the cross sections for a given DM mass $m_{\widetilde{\chi}^0_1}$ at $90\%$ confidence level from XENON1T and PandaX-II, respectively.

\item Lastly, $\chi^2_{GCE}$ in Eq.(\ref{eq.X2_total}) is constructed according to Eq.(5.1) of \cite{Calore:2014xka}:
{\small
\begin{eqnarray}
  \chi^2(\mathcal{A}) = \sum\limits_{i,j=1}^{24} \left(\frac{d\bar{N}}{dE_i}
  -\frac{dN_0}{dE_i}\right)
  \Sigma^{-1}_{ij}\left(\frac{d\bar{N}}{dE_j}   -\frac{dN_0}{dE_j}\right),
  \label{eq:X2_GCE}
\end{eqnarray} }
where $d\bar{N}/dE_i=\mathcal{A} dN/dE_i$ and $dN_0/dE_i$ are the gamma-ray spectrum predicted by NMSSM samples with a scaling factor $\mathcal{A}$ and GCE spectrum extracted from Fermi-LAT data after background modeling, respectively, in the $i$-th gamma-ray energy bin. $\mathcal{A}$ is a tuning factor accommodating the Galactic DM halo profile uncertainties which scales the theoretical predictions of $dN/dE_i$.
We define $\chi^2_{GCE}$ as the minimum value of  $\chi^2(\mathcal{A})$ with varying $\mathcal{A}$ \cite{Cao:2015loa},
\begin{eqnarray}
  \chi^2_{GCE} = min(\chi^2(\mathcal{A})), \quad \mathcal{A} \in (0.2,5).
\end{eqnarray}
$\chi^2(\mathcal{A})$ can be understood as a generalization of the 1-dimensional (1-D) $\chi^2$ in Eq.(\ref{chi2-1D}) to the 2-D case with $\sigma_i^2 \to \Sigma_{ij}^{-1}$, where $\Sigma^{-1}_{ij}$ is the inverse of a covariance matrix $\Sigma_{ij}$ which reflects the correlations among different energy bins,
defined as \cite{Calore:2014xka}
\begin{eqnarray}
\Sigma_{ij} = \left\langle \frac{dN_0}{dE_i}\frac{dN_0}{dE_j} \right\rangle -     \left\langle \frac{dN_0}{dE_i}\right\rangle\left\langle\frac{dN_0}{dE_j}    \right\rangle,
\end{eqnarray}
where the average runs over the 22 test region of interests (ROIs) in \cite{Calore:2014xka}. With $\chi^2_{GCE}$ constructed in Eq.(\ref{eq:X2_GCE}), all terms in Eq.(\ref{eq.X2_total}) are consistent in the sense of statistic definition. We refer interested readers to \cite{Calore:2014xka} for more details about $\Sigma_{ij}$.
\end{itemize}

When $\chi^2_{total}$ in the scan reaches a relatively stable minimum, we upgrade the requirements related to the observables in $\chi^2_{total}$ with hard cutting ranges from the corresponding experiments. Other constraints we further apply include the DM indirect search results from dwarf spheroidal galaxies (dSph) and the bounds from SUSY searches at LEP and LHC. More specifically, we consider:
\begin{enumerate}
  \label{constraints}
  \item The SM Higgs data \cite{ATLAS:2012tfa, CMS:2012xdj, Aad:2014aba}. We require NMSSM to accommodate a CP-even Higgs boson whose mass is near 125~GeV while decays and couplings satisfy the observed results by ATLAS and CMS. We implement this requirement by utilizing the packages HiggsBounds-5.0.0 \cite{Bechtle:2015pma} and HiggsSignal-2.0.0 \cite{Bechtle:2014ewa}. The Higgs mass we use is  $m_{h_{SM}}=(125.36\pm0.41\pm2.0)$ GeV \cite{Aad:2014aba}, indicating in order the experimental central value $\mu_i$ in Eq.(\ref{chi2-1D}), the experimental uncertainty $\sigma_{i,the}$ at $1\sigma$ and the theoretical uncertainty $\sigma_{i,exp}$ at $1\sigma$. Thus the total uncertainty in Eq.(\ref{chi2-1D}) is $\sigma_i=\sqrt{\sigma_{i,the}^2+\sigma_{i,exp}^2}$, which also applies to the following observables with similar format of data.
  \item $B$-physics constraints, i.e. $BR(B_s \rightarrow X_s\gamma)=(3.43\pm0.22\pm0.24)*10^{-4}$ \cite{Amhis:2014hma} and $BR(B_s\rightarrow \mu^+\mu^-)=(2.9\pm0.7\pm0.29)*10^{-9}$ \cite{CMS:2014xfa}, with the same form as $m_{h_{SM}}$.
  \item Constraints from SUSY searches at the LEP and LHC, which have been encoded in the package NMSSMTools-5.0, such as the mass limits on SUSY particles.
  \item Bounds on DM annihilations from Fermi-LAT gamma-ray observations of Milky Way dSphs \cite{Ackermann:2015zua} in which likelihood analysis codes are provided for constraining theoretical models\footnote{\url{http://www-glast.stanford.edu/pub_data/1048/}.}. We input the predicted gamma-ray spectrum and DM annihilation cross sections of our NMSSM samples and follow the code analysis. We apply the selection cut $\chi^2_{dSph}<2.71/2$ suggested by the codes to pick out the NMSSM samples passing the dSph constraints.
  \item Observed DM relic density $\Omega_{\widetilde{\chi}^0_1} h^2 = 0.1197* (1\pm 10\%)$ \cite{Ade:2015xua}, where the $10\%$ is included to accommodate the uncertainty of numerical calculations in MicrOMEGAs \cite{Barducci:2016pcb}.
  \item For the GCE spectrum comparison between the theoretical and experimental results, we impose the final selection criterion $\chi^2_{GCE}<35.2$. This value corresponds to the explanation of GCE by our NMSSM samples at $95\%$ confidence level with $N_{exp.}-N_{the.}=24-1$ degree of freedom (d.o.f) \cite{Cao:2015loa}, where $N_{exp.}=24$ is the number of GCE energy bins in the analysis of Fermi-LAT data \cite{Calore:2014xka} and $N_{the.}=1$ is the theoretical scaling factor $\mathcal{A}$ in Eq.(\ref{eq:X2_GCE}).
\end{enumerate}

Note that the data from DM direct detection experiments XENON1T and PandaX-II are included in Eq.(\ref{eq.X2_total}) for the $\chi^2_{total}$ scan to explore NMSSM parameter space with generally small scattering rates $\sigma^{SI}_{\widetilde{\chi}^0_1-P}$ and $\sigma^{SD}_{\widetilde{\chi}^0_1-N}$, but $\chi^2_{\sigma^{SI}_{\widetilde{\chi}^0_1-P}}$ and $\chi^2_{\sigma^{SD}_{\widetilde{\chi}^0_1-N}}$ are not necessarily equivalent to a selection cut to be under the exclusion bounds. As we mentioned in the beginning of this subsection, samples surviving all the above selection rules which can explain the GCE may not satisfy the bounds of XENON1T and PandaX-II. We name the samples obtained at this stage the \textit{GCE-samples}. In the next section, we will first analyze the characteristics of the GCE-samples. Then we confront them with the latest constraints from XENON1T and PandaX-II.

\section{Results from the scan}
\label{sec.result}

\begin{figure}[ht]
\centering
     \includegraphics[width=9cm]{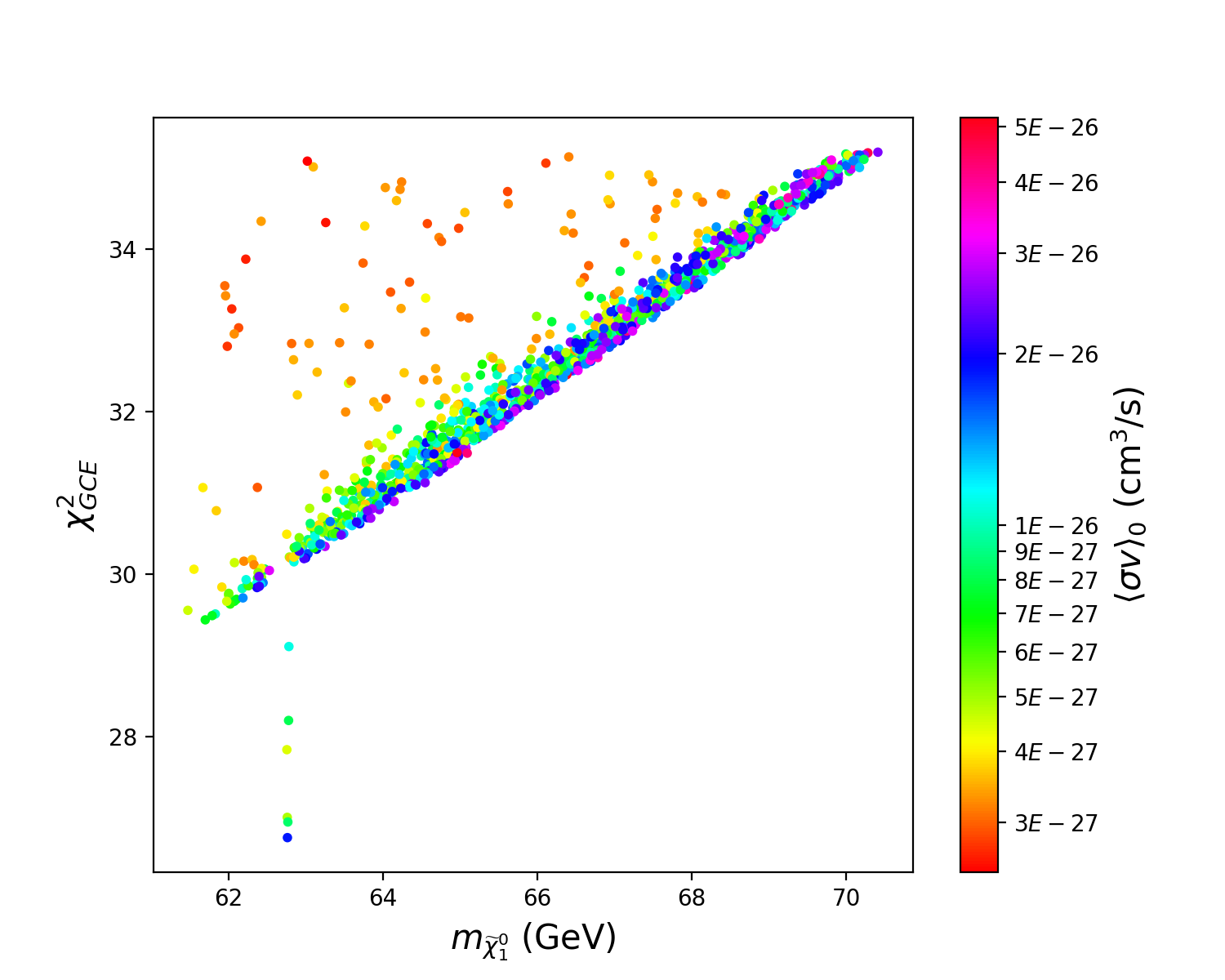}
\caption{GCE-samples plotted on the plane of $m_{\widetilde{\chi}^0_1}$ versus $\chi^2_{GCE}$ with color denoting $\langle\sigma v\rangle_0$, i.e. the DM annihilation cross section in today's Universe.}
\label{GCE_m}
\end{figure}

In this section we first present the main features of the GCE-samples obtained in Section \ref{sec.strategies}. Then we discuss their compatibility with the latest DM direct detection bounds from XENON1T and PandaX-II.

\subsection{Features of the GCE-samples}
\label{section-GCE-feature}

In Figure \ref{GCE_m}, all GCE-samples are plotted on the plane of $m_{\widetilde{\chi}^0_1}$ versus $\chi^2_{GCE}$ with color denoting $\langle\sigma v\rangle_0$, i.e. the DM annihilation cross section in today's Universe. From this figure, we found that:
\begin{itemize}
 \item DM masses $m_{\widetilde{\chi}^0_1}$ of the GCE-samples mostly locate between 60 and 70 GeV with Singlino component dominated, and a resonant effect near $m_{\widetilde{\chi}^0_1}\approx m_{h_{SM}}/2$ can be clearly identified. Since $\widetilde{\chi}^0_1 \widetilde{\chi}^0_1\to t\bar{t},W^+W^-,ZZ,hh...$ are not accessible for DM so light in today's Universe, we checked that the dominant DM annihilation channel is $\widetilde{\chi}^0_1 \widetilde{\chi}^0_1 \to b\bar{b}$.
However, we checked that some samples with $\widetilde{\chi}^0_1 \widetilde{\chi}^0_1 \to W^+W^-$ also passed various requirements in Section \ref{sec.strategies} but failed to satisfy GCE fitting criterion $\chi^2_{GCE}<35.2$, which has been noticed and pointed out in our previous work \cite{Cao:2015loa} with a different construction of $\chi^2_{total}$. The difficulty of $W^+W^-$ channel to fit the GCE comes from the over-boosted gamma-ray spectrum given the relatively large $m_W$.
  \item Generally speaking, for a given DM mass $m_{\widetilde{\chi}^0_1}$, larger today's DM annihilation cross sections $\langle\sigma v\rangle_0$ correspond to smaller $\chi^2_{GCE}$, i.e. better compatibility of the predicted gamma-ray spectrum with the observed GCE, except for the strong resonant region $m_{\widetilde{\chi}^0_1}\approx m_{h_{SM}}/2$. Understandably, a sufficiently strong production rate of the gamma-ray flux from DM annihilations is helpful to act as an additional gamma-ray source to explain the GCE.
  \item Resonance enhancement near $m_{\widetilde{\chi}^0_1}\approx m_{h_{SM}}/2$ alleviates the pressure on the NMSSM parameters of producing large couplings in $\widetilde{\chi}^0_1 \widetilde{\chi}^0_1 \to b\bar{b}$, e.g. $C_{A_i \widetilde{\chi}_1^0 \widetilde{\chi}_1^0}$ and $C_{A_i b\bar{b}}$ in Eq.(\ref{eq.bbar}). Therefore this mass region can provide the NMSSM parameters more flexibility to fine tune the shape of the predicted gamma-ray spectrum to better fit to the GCE and result in smaller $\chi^2_{GCE}$.
  \item Generically, smaller DM masses $m_{\widetilde{\chi}^0_1}$ tend to produce smaller $\chi^2_{GCE}$, since the larger DM number density with smaller DM mass can provide higher probability for DM to find each other and annihilate, yielding stronger gamma-ray flux.
\end{itemize}

\begin{figure}[ht]
  \centering
  \includegraphics[width=9cm]{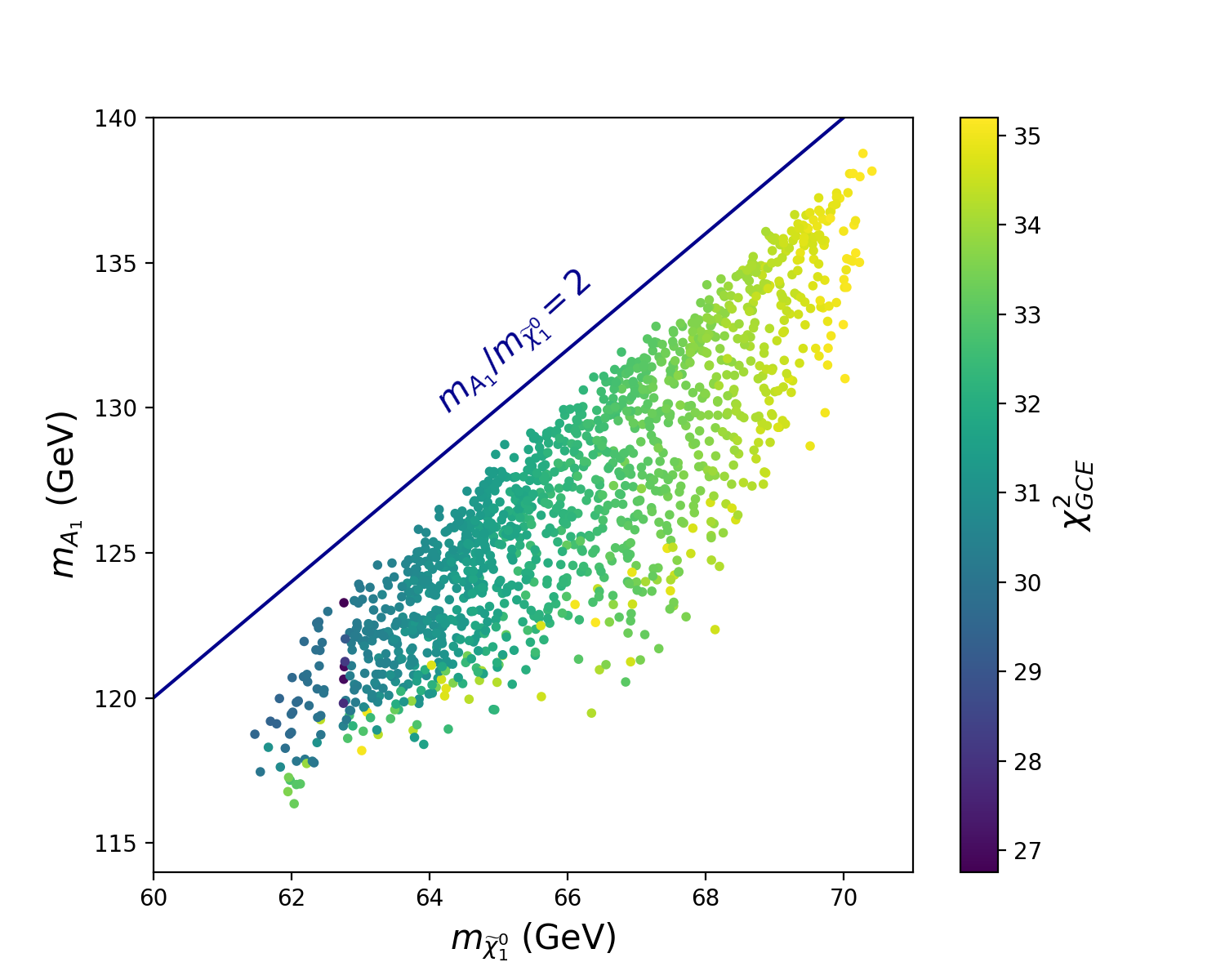}
  \caption{GCE-samples plotted on the plane of $m_{\widetilde{\chi}^0_1}$ versus $m_{A_1}$ with $\chi^2_{GCE}$ represented by colors, where a solid line indicating $m_{A_1}/m_{\widetilde{\chi}^0_1}=2$ is provided.}
  \label{mA1_mN1}
\end{figure}

Note that the DM annihilation processes in Eq.(\ref{eq:channel-general}) can also contribute to the DM annihilations in the early Universe which make DM freeze out from the SM thermal plasma. In Figure \ref{mA1_mN1} we project the GCE-samples on the plane of $m_{\widetilde{\chi}^0_1}$ versus $m_{A_1}$ with $\chi^2_{GCE}$ represented by colors, where a solid line indicating $m_{A_1}/m_{\widetilde{\chi}^0_1}=2$ is provided.
We can see that $2m_{\widetilde{\chi}^0_1}/m_{A_1}$ is near the resonant region for most samples, implying that the light CP-odd Higgs $A_1$ also play an important role in the $s$-channel annihilation in the early Universe, except in some other strong resonant scenarios such as $m_{\widetilde{\chi}^0_1}\approx m_{h_{SM}}/2$. We note that $s$-channel DM annihilations can also proceed through $Z$ boson and the CP-even Higgs bosons $H_i$. However, we checked that their contributions are small because the coupling $C_{Z \widetilde{\chi}^0_1 \widetilde{\chi}^0_1}$ is week for a Singlino-like $\widetilde{\chi}^0_1$ in the DM annihilations, and the mass relation $2 m_{\widetilde{\chi}^0_1}/m_{H_1} \gtrsim \mathcal{O}(2)$ is mostly off-resonant for the $H_1$-mediation.

\subsection{GCE confronting DM direct detection in NMSSM}
\label{section-GCE-DD}

\begin{figure}
  \centering
  \includegraphics[width=9cm]{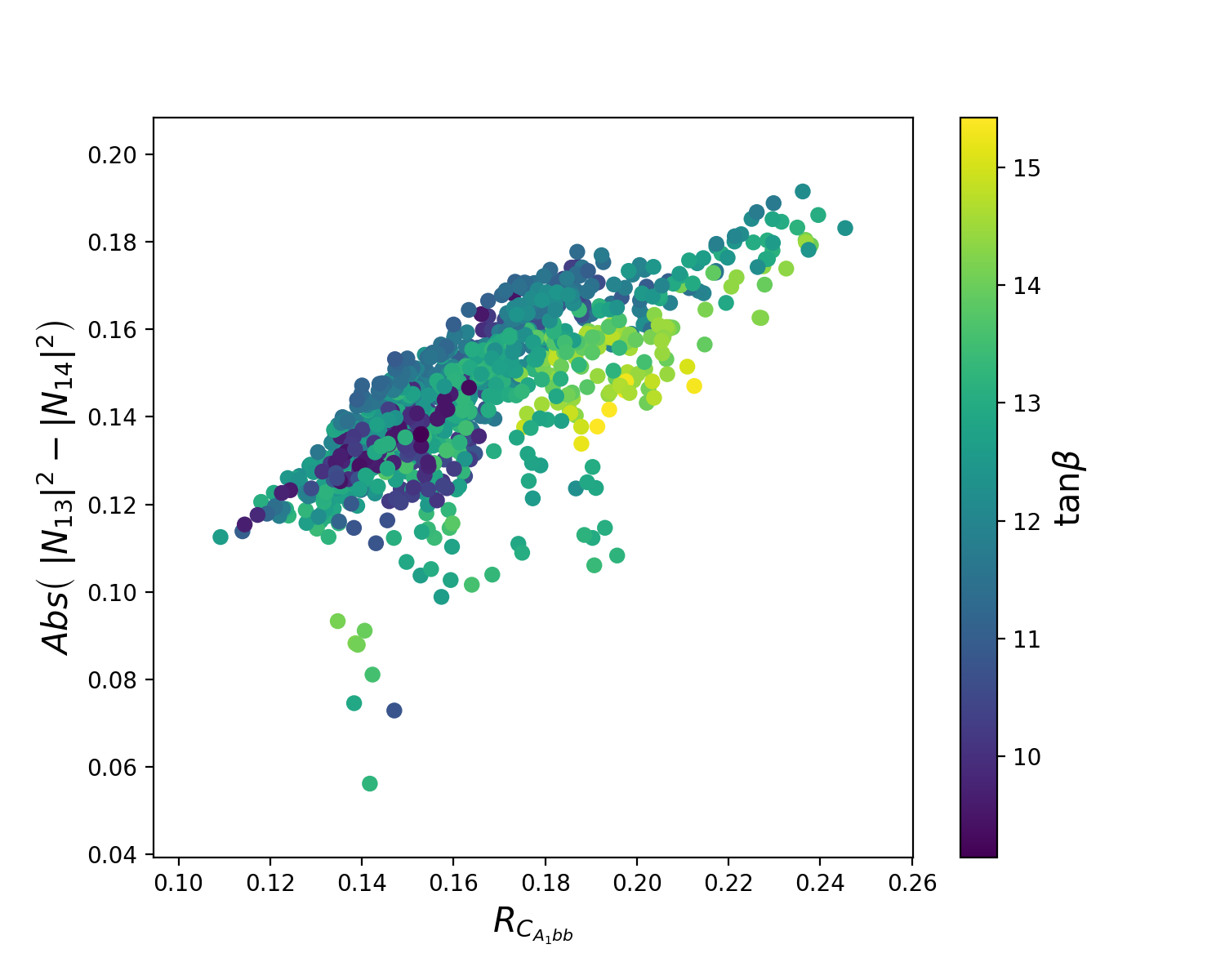}\\
  \caption{GCE-samples on the plain of $R_{C_{A_1 b\bar{b}}}\equiv C_{A_1b\bar{b}}/y_b$ versus $Abs \Big( |N_{13}|^2 - |N_{14}|^2 \Big)$ with color denoting the value of $\tan\beta$ and $y_b$ being the bottom quark Yukawa coupling in the SM.}
  \label{higgsino_Rabb_tanbeta}
\end{figure}

In Figure \ref{higgsino_Rabb_tanbeta} we plot all GCE-samples on the plain of $R_{C_{A_1 b\bar{b}}}$ versus $Abs \Big( |N_{13}|^2 - |N_{14}|^2 \Big)$ with color denoting the value of $\tan\beta$. One can clearly see the correlation pattern in terms of the increasing $\tan\beta$ as indicated in Eq.(\ref{eq:correlation}), except for some samples corresponding to the strong resonant regions in DM annihilations. We also note that the non-zero Bino and Wino components in $\widetilde{\chi}^0_1$ can obscure the correlation, but the approximation holds fairly well for a large portion of the GCE-samples.

\begin{figure}
  \centering
  \includegraphics[width=9cm]{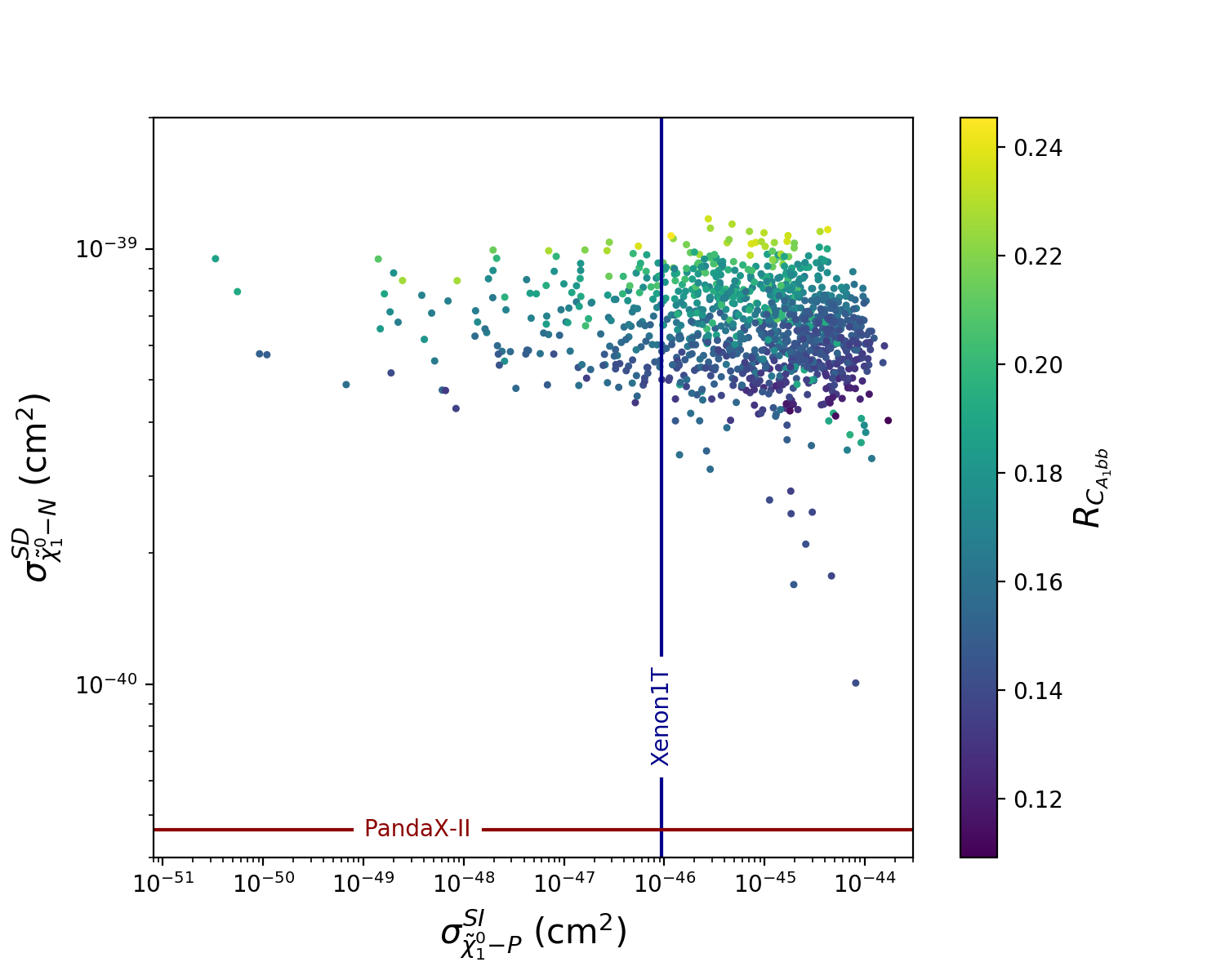}\\
  \caption{GCE-samples plotted on the plane of $\sigma^{SI}_{\widetilde{\chi}^0_1-P}$ versus $\sigma^{SD}_{\widetilde{\chi}^0_1-N}$ with color denoting the reduced coupling $R_{C_{A_1 b\bar{b}}}\equiv  C_{A_1b\bar{b}}/y_b$. Dark blue (red) line represents $\sigma^{SI}_{\widetilde{\chi}^0_1-P}=9.8\times10^{-47}\ \rm{cm^2}$ ($\sigma^{SD}_{\widetilde{\chi}^0_1-N}=4.6\times10^{-41}\ \rm{cm^2}$) limit from XENON1T (Pandax-II) at $90\%$ confidence level near $m_{\widetilde{\chi}^0_1}\simeq 62\ \rm{GeV}$.}
  \label{Nsd_Psi}
\end{figure}

Finally, in Figure \ref{Nsd_Psi} we confront the GCE-samples with the updated limits from the DM direct detections of XENON1T and PandaX-II. We project the GCE-samples on the plane of $\sigma^{SI}_{\widetilde{\chi}^0_1-P}$ versus $\sigma^{SD}_{\widetilde{\chi}^0_1-N}$ with color denoting the reduced coupling $R_{C_{A_1 b\bar{b}}}\equiv  C_{A_1b\bar{b}}/y_b$ in Eq.(\ref{eq:CA1bb}) included in the DM annihilation cross sections to explain the GCE. Since the bounds on $\sigma^{SI}_{\widetilde{\chi}^0_1-P}$ and $\sigma^{SD}_{\widetilde{\chi}^0_1-N}$ don't change much for the narrow DM mass ranges between 60 GeV and 70 GeV indicated by Figure \ref{GCE_m}, we choose their respective strongest limits at $90\%$ confidence level near $m_{\widetilde{\chi}^0_1}\simeq 62\ \rm{GeV}$, i.e. $\sigma^{SD}_{\widetilde{\chi}^0_1-N}=4.6\times10^{-41}\ \rm{cm^2}$ and $\sigma^{SI}_{\widetilde{\chi}^0_1-P}=9.8\times10^{-47}\ \rm{cm^2}$. As we can see from Figure \ref{Nsd_Psi}, all GCE-samples we obtain are excluded by $\sigma^{SD}_{\widetilde{\chi}^0_1-N}$ from PandaX-II but a small portion can still pass the $\sigma^{SI}_{\widetilde{\chi}^0_1-P}$ constraints from XENON1T. The colors representing $R_{C_{A_1 b\bar{b}}}$ clearly show that large $C_{A_1 b\bar{b}}$ which can enhance $\langle \sigma_{b\bar{b}} v \rangle _{0}$ for GCE explanation generally produce large scattering rate $\sigma^{SD}_{\widetilde{\chi}^0_1-N}$, except for some strong resonant regions.

We note that other refined scan strategy may explore more difficult corners of NMSSM parameter space providing finely tuned resonant mechanisms, which may reconcile the GCE interpretation and the DM direct detection results. However, we believe that the approximated correlation in Eq.(\ref{eq:correlation}) and the GCE-samples we obtain in Figure \ref{higgsino_Rabb_tanbeta} and in Figure \ref{Nsd_Psi} can serve as an illustration of the tension between these two DM phenomenologies in the $Z_3$-NMSSM when the singlet superfield $\hat{S}$ play an important role in the DM and Higgs sector. The future update of PandaX-nT and XENONnT experiments will continue to challenge the above scenario.

\section{Summary}
\label{sec.summary}

In this work we consider an interesting scenario
in the $Z_3$-NMSSM where the singlet $S$ and Singlino $\widetilde{S}^0$ components play important roles in the Higgs and DM sector. Guided by the analytical argument, we perform a sophisticated scan by considering various observables such as the SM Higgs data and the $B$-physics observables .
We first collect samples without applying the strict DM direct detection (DD) bounds and analyze their features about the GCE interpretation. We find that $\widetilde{\chi}^0_1$ DM are mostly Singlino-like and annihilation products are mostly the bottom quark pairs $\bar{b}b$ through a light singlet-like CP-odd Higgs $A_1$. Moreover, a good fit to the GCE spectrum generically requires sizable DM annihilation rates $\langle \sigma_{b\bar{b}} v \rangle _{0}$ in today's Universe. However, the correlation between the coupling $C_{A_1 b\bar{b}}$ in $\langle \sigma_{b\bar{b}} v \rangle _{0}$ and the coupling $C_{Z \widetilde{\chi}^0_1 \widetilde{\chi}^0_1}$ in DM-neutron Spin Dependent (SD) scattering rate $\sigma^{SD}_{\widetilde{\chi}^0_1-N}$ makes all samples we obtain for GCE explanation excluded by the PandaX-II results. Although the DM resonant annihilation scenarios may be beyond the reach of our analytical approximations and scan strategy, the aforementioned correlation can be a reasonable motivation for future experiments such as PandaX-nT to further test the NMSSM interpretation of GCE.

\section{Acknowledgement}
We thank Junjie Cao and Yang Zhang for helpful discussions, as well as Peiwen Wu for advices on the analytical arguments. This work
was supported in part by the National Natural Science Foundation of
China (NNSFC) under grant No. 11705048.

%
\bibliographystyle{unsrt}
\bibliography{GCE_DD}
%

\end{document}